\documentclass[aps,prx,twocolumn,amsmath,amssymb,superscriptaddress,letterpaper,longbibliography,floatfix,nofootinbib]{revtex4-1}
\usepackage{graphicx}
\usepackage{amsmath}
\usepackage{amssymb}
\usepackage{dsfont}
\usepackage{comment}
\usepackage{placeins}
\usepackage[colorlinks]{hyperref}
\usepackage{hypcap}
\usepackage{color}

\newcommand{\eq}[1]{Eq.~\hyperref[eq:#1]{(\ref*{eq:#1})}}
\renewcommand{\sec}[1]{\hyperref[sec:#1]{Section~\ref*{sec:#1}}}
\newcommand{\app}[1]{\hyperref[app:#1]{Appendix~\ref*{app:#1}}}
\newcommand{\tab}[1]{\hyperref[tab:#1]{Table~\ref*{tab:#1}}}
\newcommand{\fig}[1]{\hyperref[fig:#1]{Figure~\ref*{fig:#1}}}
\newcommand{\figa}[2]{\hyperref[fig:#1]{Figure~\ref*{fig:#1}#2}}
\newcommand{\figx}[2]{\hyperref[fig:#1]{Figure~\ref*{fig:#1}(#2)}}
\newcommand{\thm}[1]{\hyperref[thm:#1]{Theorem~\ref*{thm:#1}}}
\newcommand{\lem}[1]{\hyperref[lem:#1]{Lemma~\ref*{lem:#1}}}
\newcommand{\cor}[1]{\hyperref[cor:#1]{Corollary~\ref*{cor:#1}}}
\newcommand{\defn}[1]{\hyperref[def:#1]{Definition~\ref*{def:#1}}}
\newcommand{\alg}[1]{\hyperref[alg:#1]{Algorithm~\ref*{alg:#1}}}

\def\bra#1{\mathinner{\langle{#1}|}}
\def\ket#1{\mathinner{|{#1}\rangle}}
\newcommand{\braket}[2]{\langle #1|#2\rangle}

\newcommand{\ignore}[1]{}

\newcommand{\googsb}{\affiliation{Google Inc., Santa Barbara, CA 93117, USA}}
\newcommand{\googla}{\affiliation{Google Inc., Venice, CA 90291, USA}}
\newcommand{\harvard}{\affiliation{Department of Chemistry, Harvard University, Cambridge, MA 02138, USA}}
\newcommand{\tufts}{\affiliation{Department of Physics, Tufts University, Medford, MA 02155, USA}}
\newcommand{\lbnl}{\affiliation{Computational Research Division, Lawrence Berkeley National Laboratory, Berkeley CA 94720, USA}}
\newcommand{\ucsb}{\affiliation{Department of Physics, University of California, Santa
Barbara, CA 93106, USA}}
\newcommand{\ucl}{\affiliation{Center for Computational Science and Department of Chemistry, University College London, WC1H 0AJ, UK}}

\input{Qcircuit}
\begin{document}

\title{Scalable Quantum Simulation of Molecular Energies}

\author{P. J. J. O'Malley}
\thanks{These authors contributed equally to this work.}
\ucsb
\author{R. Babbush}
\thanks{These authors contributed equally to this work.}
\googla
\author{I. D. Kivlichan}
\harvard
\author{J. Romero}
\harvard
\author{J. R. McClean}
\lbnl
\author{R. Barends}
\googsb
\author{J. Kelly}
\googsb
\author{P. Roushan}
\googsb
\author{A. Tranter}
\tufts
\ucl
\author{N. Ding}
\googla
\author{B. Campbell}
\ucsb
\author{Y. Chen}
\googsb
\author{Z. Chen}
\ucsb
\author{B. Chiaro}
\ucsb
\author{A. Dunsworth}
\ucsb
\author{A. G. Fowler}
\googsb
\author{E. Jeffrey}
\googsb
\author{E. Lucero}
\googsb
\author{A. Megrant}
\googsb
\author{J. Y. Mutus}
\googsb
\author{M. Neeley}
\googsb
\author{C. Neill}
\ucsb
\author{C. Quintana}
\ucsb
\author{D. Sank}
\googsb
\author{A. Vainsencher}
\ucsb
\author{J. Wenner}
\ucsb
\author{T. C. White}
\googsb
\author{P. V. Coveney}
\ucl
\author{P. J. Love}
\tufts
\author{H. Neven}
\googla
\author{A. Aspuru-Guzik}
\harvard
\author{J. M. Martinis}
\googsb \ucsb

\begin{abstract}
\vspace{.1cm}
 We report the first electronic structure calculation performed on a quantum computer without exponentially costly precompilation.
 We use a programmable array of superconducting qubits to compute the energy surface of molecular hydrogen using two distinct quantum algorithms.
 First, we experimentally execute the unitary coupled cluster method using the variational quantum eigensolver.
 Our efficient implementation predicts the correct dissociation energy to within chemical accuracy of the numerically exact result.
 Second, we experimentally demonstrate the canonical quantum algorithm for chemistry, which consists of Trotterization and quantum phase estimation.
 We compare the experimental performance of these approaches to show clear evidence that the variational quantum eigensolver is robust to certain errors.
 This error tolerance inspires hope that variational quantum simulations of classically intractable molecules may be viable in the near future. 
\end{abstract}

\maketitle

Universal and efficient simulation of physical systems \cite{Lloyd1996} is among the most compelling applications of quantum computing. In particular, quantum simulation of molecular energies \cite{Aspuru-Guzik2005}, which enables numerically exact prediction of chemical reaction rates, promises significant advances in our understanding of chemistry and could enable \emph{in silico} design of new catalysts, pharmaceuticals and materials. As scalable quantum hardware becomes increasingly viable \cite{Barends2014,Kelly2015,Corcoles2015,Riste2015,Barends2015}, chemistry simulation has attracted significant attention \cite{Whitfield2010,Kassal2010,Jones2012,Wecker2014,Poulin2014,BabbushTrotter,Whitfield2013b,Veis2014,Veis2015,Tranter2015,BabbushAQChem,Peruzzo2013,Yung2013,McClean2014,Wecker2015,McClean2015,Toloui2013,BabbushSparse1,BabbushSparse2,Trout2015,Moll2015} since classically intractable molecules require a relatively modest number of qubits and because solutions have commercial value associated with their chemical applications \cite{Mueck2015}.

The fundamental challenge in building a quantum computer is realizing high-fidelity operations in a scalable architecture \cite{Martinis2015}. Superconducting qubits have made rapid progress in recent years \cite{Barends2014,Kelly2015,Corcoles2015,Riste2015} and can be fabricated in microchip foundries and manufactured at scale \cite{Johnson2011}. Recent experiments have shown logic gate fidelities at the threshold required for quantum error correction \cite{Barends2014} and dynamical suppression of bit-flip errors \cite{Kelly2015}. Here, we use the device reported in \cite{Kelly2015,Barends2015,Barends2016} to implement and compare two quantum algorithms for chemistry. We have previously characterized our hardware using randomized benchmarking \cite{Kelly2015} but related metrics (e.g. fidelities) only loosely bound how well our devices can simulate molecular energies. Thus, studying the performance of hardware on small instances of real problems is an important way to measure progress towards viable quantum computing. 

Our first experiment demonstrates the recently-proposed variational quantum eigensolver (VQE), introduced in \cite{Peruzzo2013}. Our VQE experiment achieves chemical accuracy and is the first scalable quantum simulation of molecular energies performed on quantum hardware, in the sense that our algorithm is efficient and does not benefit from exponentially costly precompilation \cite{Smolin2013a}. When implemented using a unitary coupled cluster ansatz, VQE cannot be efficiently simulated classically and empirical evidence suggests that answers are accurate enough to predict chemical rates \cite{Peruzzo2013,Yung2013,Wecker2015,McClean2014,McClean2015}. Because VQE only requires short state preparation and measurement sequences, it has been suggested that classically intractable computations might be possible using VQE without the overhead of error correction \cite{McClean2015,Wecker2015}. Our experiments substantiate this notion; the robustness of the VQE to systematic device errors allows the experiment to achieve chemical accuracy.

\begin{figure*}[t]
\includegraphics[width=\linewidth]{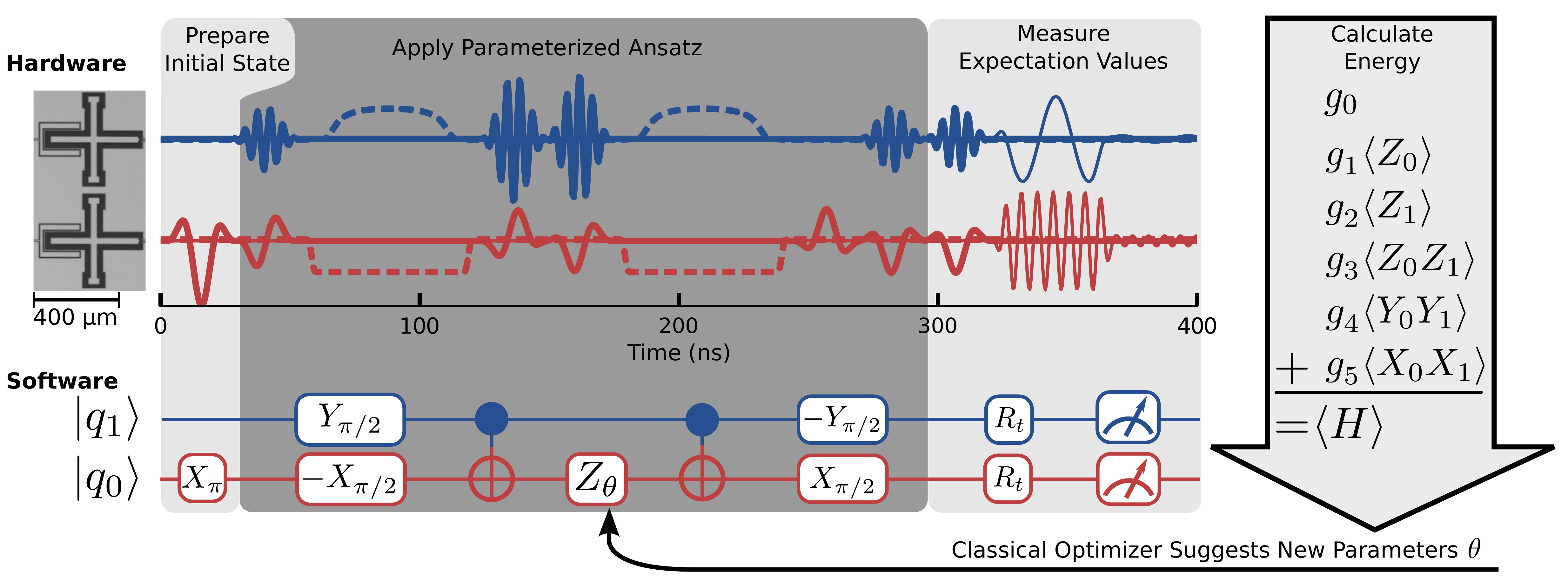}
\caption{\label{fig:vqe} \textbf{Hardware and software schematic of the variational quantum eigensolver.} (Hardware) micrograph shows two Xmon transmon qubits and microwave pulse sequences to perform single-qubit rotations (thick lines), DC pulses for two-qubit entangling gates (dashed lines), and microwave spectrosopy tones for qubit measurements (thin lines). (Software) quantum circuit diagram shows preparation of Hartree-Fock state, followed by application of the unitary coupled cluster ansatz in \eq{xy} and efficient partial tomography ($R_t$) to measure the expectation values in \eq{H2}. Finally, the total energy is computed according to \eq{sum} and provided to a classical optimizer which suggests new parameters.}
\end{figure*}

Our second experiment realizes the original algorithm for the quantum simulation of chemistry, introduced in \cite{Aspuru-Guzik2005}. This approach involves Trotterized simulation \cite{Trotter1959} and the quantum phase estimation algorithm (PEA) \cite{Kitaev1995}. We experimentally perform this entire algorithm, including both key components, for the first time. While PEA has asymptotically better scaling in terms of precision than VQE, long and coherent gate sequences are required for its accurate implementation.

The phase estimation component of the canonical quantum chemistry algorithm has been demonstrated in a photonic system \cite{Lanyon2010}, a nuclear magnetic resonance system \cite{Du2010}, and a nitrogen-vacancy center system \cite{Wang2014}. While all three experiments obtained molecular energies to incredibly high precision, none of the experiments implemented the propagator in a scalable fashion (e.g. using Trotterization) as doing so requires long coherent evolutions. Furthermore, none of these experiments used more than a single qubit or qutrit to represent the entire molecule. This was possible due to the use of the configuration basis, which is not scalable but renders the experimental challenge much easier. Furthermore, all of these implementations applied the logic gates with a single, totally controlled pulse, as opposed to compiling the algorithm to a universal set of gates as we do.

There have been two previous experimental demonstrations of VQE: first in a photonic system \cite{Peruzzo2013} and later in an ion trap \cite{Shen2015}. Both experiments validated the variational approach and the latter implemented an ansatz based on unitary coupled cluster. All prior experiments focused on either molecular hydrogen \cite{Lanyon2010,Du2010} or helium hydride \cite{Peruzzo2013,Wang2014,Shen2015} but none of these prior experiments employed a scalable qubit representation such as second quantization. Instead, all five prior experiments represent the Hamiltonian in a configuration basis that cannot be efficiently decomposed as a sum of local Hamiltonians, and then exponentiate this exponentially large matrix as a classical preprocessing step \cite{Lanyon2010,Du2010,Wang2014,Peruzzo2013,Shen2015}.

Until this work, important aspects of scalable chemistry simulation such as the Jordan-Wigner transformation \cite{Somma2002} or the Bravyi-Kitaev transformation \cite{Bravyi2002,Seeley2012} had never been used to represent a molecule in an experiment; however,  prior experiments such as \cite{Barends2015} have previously used the Jordan-Wigner representation to simulate fermions on a lattice. In both experiments presented here, we simulate the dissociation of molecular hydrogen in the minimal basis of Hartree-Fock orbitals, represented using the Bravyi-Kitaev transformation of the second quantized molecular Hamiltonian \cite{Tranter2015}. As shown in \app{ham}, the molecular hydrogen Hamiltonian can be scalably written as
\begin{equation}
\label{eq:H2}
H = g_0 \mathds{1} + g_1 Z_0 + g_2 Z_1 + g_3 Z_0 Z_1 + g_4 Y_0 Y_1 + g_5 X_0 X_1
\end{equation}
where $\{X_i, Z_i, Y_i\}$ denote Pauli matrices acting on the $i^\textrm{th}$ qubit and the real scalars $\{g_\gamma\}$ are efficiently computable functions of the hydrogen-hydrogen bond length, $R$.

The ground state energy of \eq{H2} as a function of $R$ defines an energy surface. Such energy surfaces are used to compute chemical reaction rates which are exponentially sensitive to changes in energy. If accurate energy surfaces are obtained, one can use established methods such as classical Monte Carlo or Molecular Dynamics simulations to obtain accurate free energies, which provide the rates directly via the Erying equation \cite{Helgaker2002}. At room temperature, a relative error in energy of $1.6 \times 10^{-3}$ Hartree (1 kcal/mol or 0.043 eV) translates to a chemical rate that differs from the true value by an order of magnitude; therefore, $1.6 \times 10^{-3}$ Hartree is known as ``chemical accuracy'' \cite{Helgaker2002}.
Our goal then is to compute the lowest energy eigenvalues of \eq{H2} as a function of $R$, to within chemical accuracy.

\begin{figure*}[t]
\includegraphics[width=\linewidth]{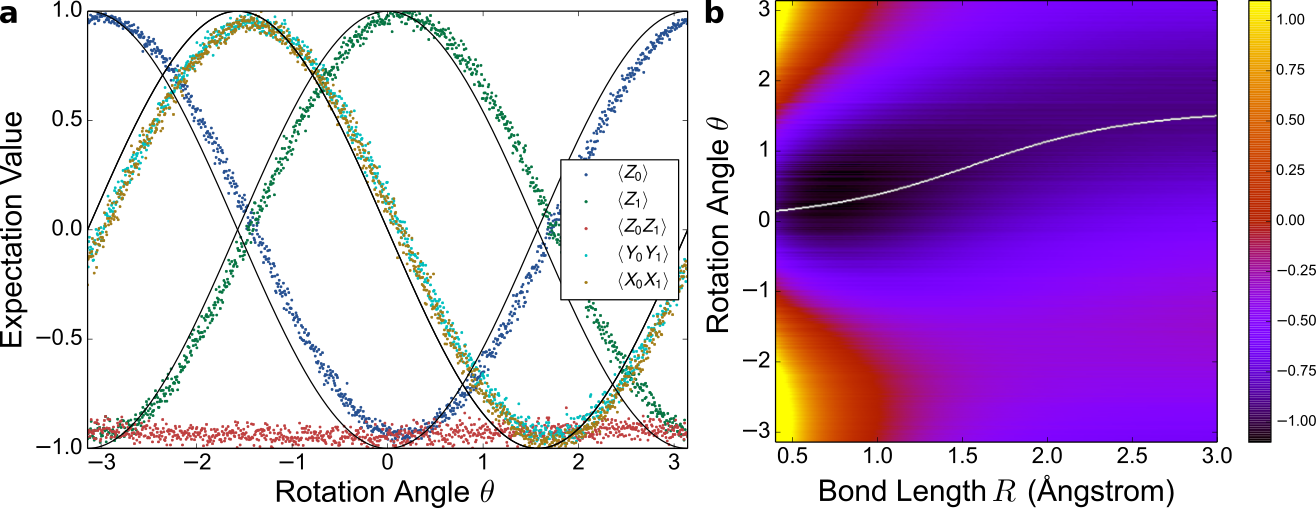}
\caption{\label{fig:data} \textbf{Variational quantum eigensolver: raw data and computed energy surface.} (\textbf{a}) Data showing the expectation values of terms in \eq{H2} as a function of $\theta$, as in \eq{xy}. Black lines nearest to the data show the theoretical values. While such systematic phase errors would prove disastrous for PEA, our VQE experiment is robust to this effect. (\textbf{b}) Experimentally measured energies (in Hartree) as a function of $\theta$ and $R$. This surface is computed from \figa{data}{a} according to \eq{sum}. The white curve traces the theoretical minimum energy; the values of theoretical and experimental minima at each $R$ are plotted in \figa{results}{a}. Errors in this surface are given in \fig{error_surface}.}
\end{figure*}

\section{Variational quantum eigensolver}

Many popular classical approximation methods for the electronic structure problem involve optimizing a parameterized guess wavefunction (known as an ``ansatz'') according to the variational principle \cite{Helgaker2002}. If we parameterize an ansatz $\ket{\varphi(\vec \theta)}$ by the vector $\vec \theta$ then the variational principle holds that 
\begin{equation}
\label{eq:var}
\frac{\bra{\varphi(\vec \theta)} H \ket{\varphi(\vec \theta)}}{\braket{\varphi(\vec \theta)}{\varphi(\vec \theta)}} \geq E_0,
\end{equation}
where $E_0$ is the smallest eigenvalue of the Hamiltonian $H$. Accordingly, $E_0$ can be estimated by selecting the parameters $\vec \theta$ which minimize the left-hand side of \eq{var}.

While the ground state wavefunction is likely to be in superposition over an exponential number of states in the basis of molecular orbitals, most classical approaches restrict the ansatz to the support of polynomially many basis elements due to memory limitations. However, quantum circuits can prepare entangled states which are not known to be efficiently representable classically. In VQE, the state $\ket{\varphi(\vec \theta)}$ is parameterized by the action of a quantum circuit $U(\vec \theta)$ on an initial state $\ket{\phi}$, i.e. $\ket{\varphi(\vec \theta)} \equiv U(\vec \theta) \ket{\phi}$. Even if $\ket{\phi}$ is a simple product state and $U(\vec \theta)$ is a very shallow circuit, $\ket{\varphi(\vec \theta)}$ can contain complex many-body correlations and span an exponential number of standard basis states.

We can express the mapping $U(\vec \theta)$ as a concatenation of parameterized quantum gates,  $U_1 (\theta_1) U_2 (\theta_2) \cdots U_n(\theta_n)$. In this work, we parameterize our circuit according to unitary coupled cluster theory \cite{Yung2013,McClean2015,Wecker2015}. As described in \app{ucc}, unitary coupled cluster predicts that the ground state of \eq{H2} can be expressed as
\begin{equation}
\label{eq:xy}
\ket{\varphi (\theta)} = e^{-i \, \theta \, X_0 Y_1} \ket{01},
\end{equation}
where $\ket{\phi} = \ket{01}$ is the Hartree-Fock (mean-field) state of molecular hydrogen in the representation of \eq{H2}. As discussed in \app{ucc}, unitary coupled cluster is widely believed to be classically intractable and is known to be strictly more powerful than the ``gold standard'' of classical electronic structure theory, coupled cluster \cite{Helgaker2002,Hoffmann1988,Bartlett1989,Taube2006}. The gate model circuit that performs this unitary mapping is shown in the software section of \fig{vqe}.

\begin{figure*}[t]
\includegraphics[width=\linewidth]{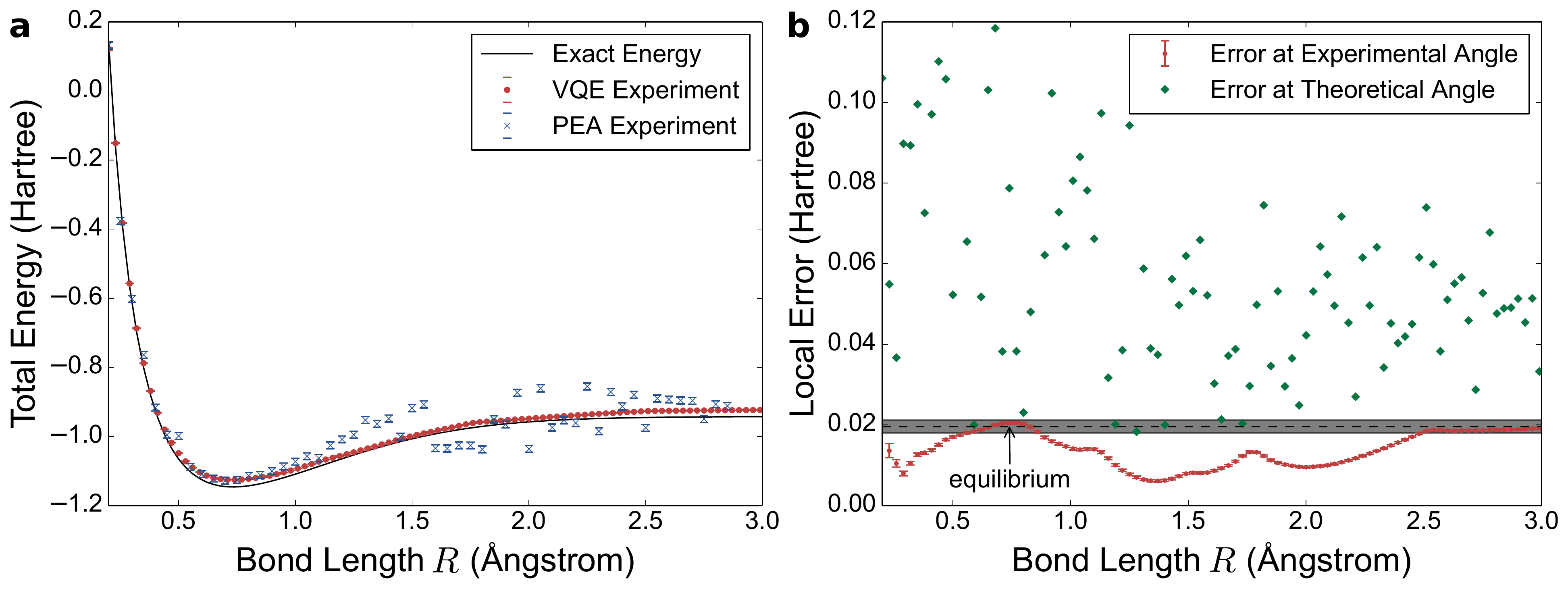}
\caption{\label{fig:results} \textbf{Computed H$_2$ energy curve and errors.} (\textbf{a}) Energy surface of molecular hydrogen as determined by both VQE and PEA. VQE approach shows dissociation energy error of $(8 \pm 5) \times 10^{-4}$ Hartree (error bars on VQE data are smaller than markers). PEA approach shows dissociation energy error of $(1 \pm 1) \times 10^{-2}$ Hartree. (\textbf{b}) Errors in VQE energy surface. Red dots show error in the experimentally determined energies. Green diamonds show the error in the energies that would have been obtained experimentally by running the circuit at the theoretically optimal $\theta$ instead of the experimentally optimal $\theta$. The discrepancy between blue and red dots provides experimental evidence for the robustness of VQE which could not have been anticipated via numerical simulations. The gray band encloses the chemically accurate region relative to the experimental energy of the atomized molecule. The dissociation energy is relative to the equilibrium geometry, which falls within this envelope.}
\end{figure*}

VQE solves for the parameter vector $\vec \theta$ with a classical optimization routine. First, one prepares an initial ansatz $\ket{\varphi(\vec \theta_0)}$ and then estimates the ansatz energy $E(\vec \theta_0)$ by measuring the expectation values of each term in \eq{H2} and summing these values together as
\begin{equation}
\label{eq:sum}
E(\vec \theta) = \sum_{\gamma} g_\gamma \bra{\varphi(\vec \theta)} H_\gamma \ket{\varphi(\vec \theta)},
\end{equation}
where the $g_\gamma$ are scalars and the $H_\gamma$ are local Hamiltonians as in \eq{H2}. The initial guess $\vec \theta_0$ and the corresponding objective value $E(\vec \theta_0)$ are then fed to a classical greedy minimization routine (e.g. gradient descent), which then suggests a new setting of the parameters $\vec \theta_1$. The energy $E(\vec \theta_1)$ is then measured and returned to the classical outer loop. This continues for $m$ iterations until the energy converges to a minimum value $E(\vec \theta_m)$ which represents the VQE approximation to $E_0$.

Because our experiment requires only a single variational parameter, as in \eq{xy}, we elected to scan a thousand different values of $\theta \in [-\pi, \pi )$ in order to obtain expectation values which define the entire potential energy curve. We did this to simplify the classical feedback routine but at the cost of needing slightly more experimental trials. These expectation values are shown in \figa{data}{a} and the corresponding energy surfaces at different bond lengths are shown in \figa{data}{b}. The energy surface in \figa{data}{b} was locally optimized at each bond length to emulate an on-the-fly implementation.

\figa{results}{a} shows the exact and experimentally determined energies of molecular hydrogen at different bond lengths. The minimum energy bond length ($R = 0.72$ \AA) corresponds to the equilibrium bond length, whereas the asymptote on the right part of the curve corresponds to dissociation into two hydrogen atoms. The energy difference between these points is the dissociation energy, and the exponential of this quantity determines the chemical dissociation rate. Our VQE experiment correctly predicts this quantity with an error of $(8 \pm 5) \times 10^{-4}$ Hartree, which is below the chemical accuracy threshold. Error bars are computed with Gaussian process regression \cite{Bishop2006} which interpolates the energy surface based on local errors from the shot-noise limited expectation value measurements in \figa{data}{a}.

Errors in our simulation as a function of $R$ are shown in \figa{results}{b}. The curve in \figa{results}{b} becomes nearly flat past $R = 2.5$ \AA $\,$ because the same angle is experimentally chosen for each $R$ past this point. Note that the experimental energies are always greater than or equal to the exact energies due to the variational principle. \figa{results}{b} shows that VQE has substantial robustness to systematic errors. While this possibility had been previously hypothesized \cite{McClean2015}, we report the first experimental signature of robustness and show that it allows for a successful computation of the dissociation energy. By performing (inefficient) classical simulations of the circuit in \fig{vqe}, we identify the theoretically optimal value of $\theta$ at each $R$. In fact, for this system, at every value of $R$ there exists $\theta$ such that $E (\theta) = E_0$. However, due to experimental error, the theoretically optimal value of $\theta$ differs substantially from the experimentally optimal value of $\theta$. This can be seen in \figa{results}{b} from the large discrepancy between the green diamonds (experimental energy errors at theoretically optimal $\theta$) and the red dots (experimental energy errors at experimentally optimal $\theta$). The experimental energy curve at theoretically optimal $\theta$ shows an error in the dissociation energy of $1.1 \times 10^{-2}$ Hartree, which is more than an order of magnitude worse. One could anticipate this discrepancy by looking at the raw data in \figa{data}{a} which shows that the experimentally measured expectation values deviate considerably from the predictions of theory. In a sense, the green diamonds in \figa{results}{b} show the performance of a non-variational algorithm, which in theory gives the exact answer, but in practice fails due to systematic errors.

\begin{figure*}[t]
\includegraphics[width=\linewidth]{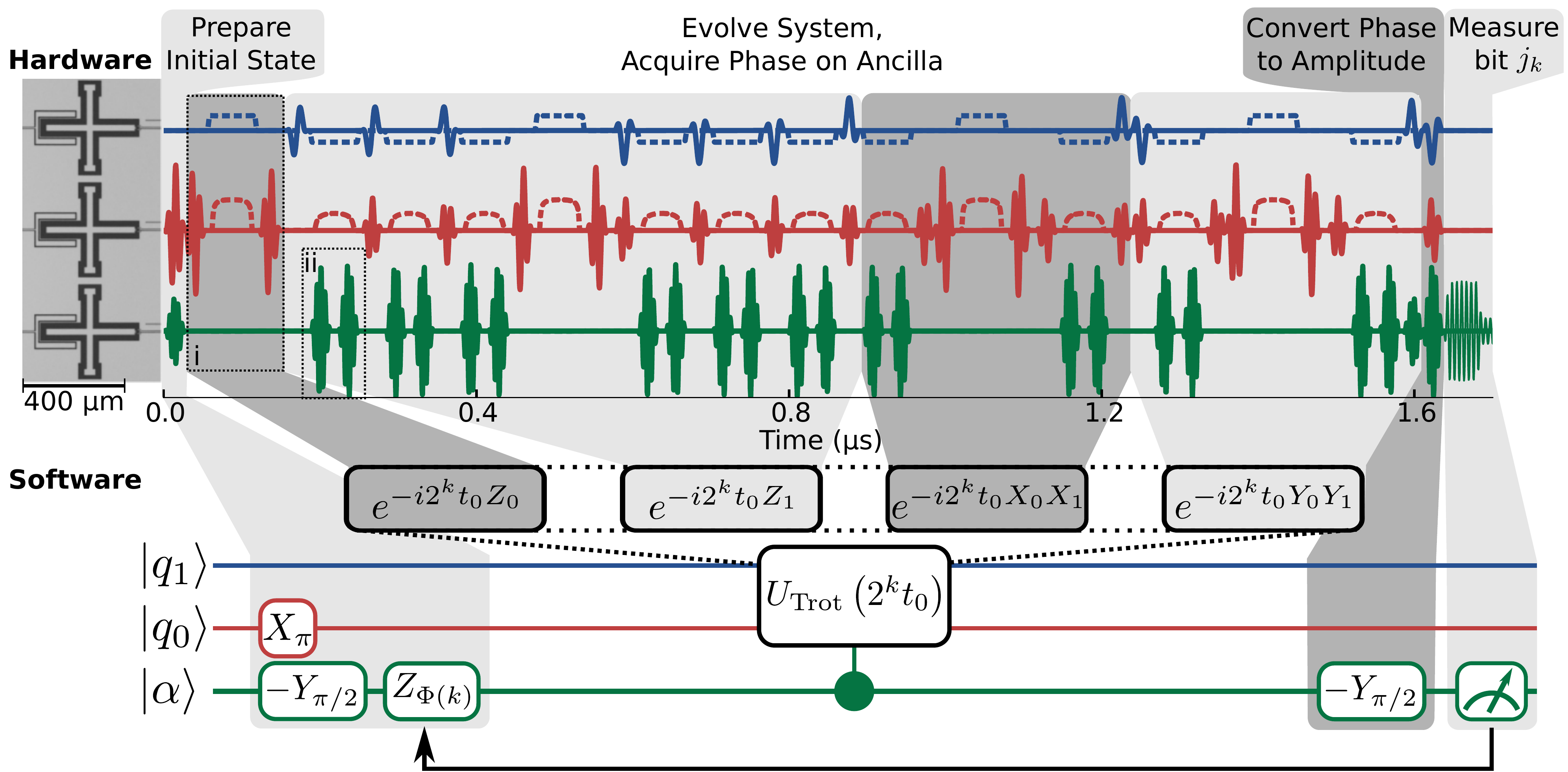}
\caption{\label{fig:pea} \textbf{Hardware and software schematic of the Trotterized phase estimation algorithm.} (Hardware) a micrograph shows three Xmon transmon qubits and microwave pulse sequences, including (\textbf{i}) the variable amplitude CZ$_\phi$ (not used in \fig{vqe}) and (\textbf{ii}) dynamical decoupling pulses not shown in logical circuit. (Software) state preparation includes putting the ancilla in a superposition state and compensating for previously measured bits of the phase using the gate $Z_{\Phi_k}$ (see text). The bulk of the circuit is the evolution of the system under a Trotterized Hamiltonian controlled by the ancilla. Bit $j_k$ is determined by a majority vote of the ancilla state over one thousand repetitions.}
\end{figure*}

\section{Phase estimation algorithm}

We also report an experimental demonstration of the original quantum algorithm for chemistry \cite{Aspuru-Guzik2005}. Similar to VQE, the first step of this algorithm is to prepare the system register in a state having good overlap with the ground state of the Hamiltonian $H$. In our case, we begin with the Hartree-Fock state, $\ket{\phi}$. We then evolve this state under $H$ using a Trotterized approximation to the time-evolution operator. The execution of this unitary is controlled on an ancilla initialized in the superposition state $(\ket{0} + \ket{1})/\sqrt{2}$. The time-evolution operator can be approximated using Trotterization \cite{Trotter1959} as
\begin{equation}
\label{eq:trot}
e^{- i H t} = e^{-i t \sum_\gamma g_\gamma H_\gamma} \approx U_\textrm{Trot}(t) \equiv \left(\prod_{\gamma} e^{-i g_\gamma H_\gamma t / \rho}\right)^\rho
\end{equation}
where the $H_\gamma$ are local Hamiltonians as in \eq{H2} and the error in this approximation depends linearly on the time step $\rho^{-1}$ \cite{Trotter1959}. Application of the propagator induces a phase on the system register so that
\begin{equation}
\label{eq:upea}
e^{- i H t} \ket{\phi} = \left(\sum_n  e^{-i E_n t} \ket{n}\!\bra{n} \right) \ket{\phi} = \sum_n a_n e^{-i E_n t} \ket{n}
\end{equation}
where $\ket{n}$ are eigenstates of the Hamiltonian such that $H \ket{n} = E_n \ket{n}$ and $a_n = \braket{n}{\phi}$. By controlling this evolution on the ancilla superposition state, one entangles the system register with the ancilla. Accordingly, by measuring the phase between the $\ket{0}$ state and $\ket{1}$ state of the ancilla, one measures the phase $E_n t$ and collapses the system register to the state $\ket{n}$ with probability $|a_n|^2$.

Our PEA implementation is based on a modification of Kitaev's iterative phase estimation algorithm \cite{Whitfield2010,Kitaev1995}. The circuit used is shown in \fig{pea} and detailed descriptions of the subroutines used to control $U_\textrm{Trot}(2^k t_0)$ on an ancilla are shown in \app{pea}. The rotation $Z_{\Phi(k)}$ in \fig{pea} feeds back classical information from the prior $k-1$ measurements using phase kickback as
\begin{equation}
\label{eq:S}
\Phi\left(k\right) = \pi \sum_{\ell = 0}^{k - 1} \frac{j_{\ell}}{2^{\ell - k + 1}}.
\end{equation}
With iterative phase estimation, one measures the phase accumulated on the system one bit at a time. Even when $a_0$ is very small, one can use iterative phase estimation to measure eigenvalues if the system register remains coherent throughout the entire phase determination. Since the Hartree-Fock state has strong overlap with the ground state of molecular hydrogen (i.e. $|\braket{0}{\phi}|^2 > 0.5$) we were able to measure each bit independently with a majority-voting scheme, reducing coherence requirements. For $b$ bits, the ground state energy is digitally computed as a binary expansion of the measurement outcomes,
\begin{equation}
E_0^b = - \frac{\pi}{t_0} \sum_{k=0}^{b-1}\frac{j_k}{2^{k+1}}.
\end{equation}

Experimentally computed energies are plotted alongside VQE results in \figa{results}{a}. Because energies are measured digitally in iterative phase estimation, the experimentally determined PEA energies in \figa{results}{a} agree exactly with theoretical simulations of \fig{pea}, which differ from the exact energies due to the approximation of \eq{trot}. The primary difficulty of the PEA experiment is that the controlled application of $U_\textrm{Trot}(2^k t_0)$ requires complex quantum circuitry and long coherent evolutions. Accordingly, we approximated the propagator in \eq{trot} using a single Trotter step ($\rho = 1$), which is not sufficient for chemical accuracy. Our PEA experiment shows an error in the dissociation energy of $(1 \pm 1) \times 10^{-2}$ Hartree.

In addition to taking only one Trotter step, we performed classical simulations of the error in \eq{trot} under different orderings of the $H_\gamma$ in order to find the optimal Trotter sequences at each value of $R$. The Trotter sequences used in our experiment as well as parameters such as $t_0$ are reported in the \app{pea}. Since this optimization is intractable for larger molecules, our PEA protocol benefited from inefficient classical preprocessing (unlike our VQE implementation). Nevertheless, this is the first time the canonical quantum algorithm for chemistry has been executed in its entirety and as such, represents a significant step towards scalable implementations.

\section{Experimental Methods}
Both algorithms are implemented with a superconducting quantum system based on the Xmon \cite{Barends2013}, a variant of the planar transmon qubit \cite{Koch2007}, in a dilution refrigerator with a base temperature of 20 mK. Each qubit consists of a SQUID (superconducting quantum interference device), which provides a tunable nonlinear inductance, and a large X-shaped capacitor; qubit frequencies are tunable up to 6 GHz and have a nonlinearity of $(\omega_{21} - \omega_{10}) = -0.22 \, \textrm{GHz}$. The qubits are capacitively coupled to their nearest neighbors in a linear chain pattern, with coupling strengths of $30 \, \textrm{MHz}$. Single-qubit quantum gates are implemented with microwave pulses and tuned using closed-loop optimization with randomized benchmarking \cite{Kelly2014}. Qubit state measurement is performed in a dispersive readout scheme with capacitively coupled resonators at 6.6-6.8 GHz \cite{Kelly2015}. For details of the device fabrication and conventions for reporting qubit parameters, see \cite{Kelly2015}.

Our entangling operation is a controlled-phase (CZ$_\phi$) gate, accomplished by holding one of the qubits at a fixed frequency while adiabatically tuning the other close to an avoided level crossing of the $\ket{11}$ and $\ket{02}$ states \cite{Barends2014}. To produce the correct phase change $\phi$, the acquired phase is measured with quantum state tomography versus the amplitude of the trajectory, and the amplitude for any given $\phi$ is then determined via interpolation \cite{Barends2015}. To minimize leakage out of the computation subspace during this operation, we increase the gate duration from the previously used 40 ns to 50 ns, and then shape the pulse trajectory. The CZ$_\phi$ gate as implemented here has a range of approximately 0.25 to 5.0 rad; for smaller values of $\phi$, parasitic interactions with other qubits become nontrivial, and for larger $\phi$, leakage is significant. For $\phi$ outside this range, the total rotation is accomplished with two physical gates. For CZ$_\phi$ gates with $\phi=\pi$, the amplitude and shape of the trajectory are further optimized with ORBIT \cite{Kelly2014}. CZ$_{\phi \ne \pi}$ is only necessary in the PEA experiment (see \app{pea}).

The gates used to implement both VQE and PEA are shown in \app{vqe} and \app{pea}. A single VQE sequence consists of 11 single-qubit gates and two CZ$_\pi$ gates. A PEA sequence has at least 51 single-qubit gates, four CZ$_{\phi \ne \pi}$ gates, and ten CZ$_\pi$ gates; more were required when not all $\phi$ values are within the range that could be performed with a single physical gate.

\section{Conclusion}

We report the use of quantum hardware to experimentally compute the potential energy curve of molecular hydrogen using both PEA and VQE. We perform the first experimental implementation of the Trotterized molecular time-evolution operator and then measure energies using PEA. Due to the costly nature of Trotterization, we are able to implement only a single Trotter step, which is not enough to achieve chemical accuracy. By contrast, our VQE experiment achieves chemical accuracy and shows significant robustness to certain types of error. The comparison of these two approaches suggests that adaptive algorithms (e.g. VQE) may generally be more resilient for pre-error corrected quantum computing than traditional gate model algorithms (e.g. PEA).

The robustness of VQE is partially a consequence of the adaptive nature of the algorithm; the classical outer loop of VQE helps to avoid systematic errors by acting similarly to the calibration loops used to tune individual quantum gates \cite{Kelly2014}. This minimization procedure treats the energy functional as a black box in that no assumptions are made about the actual circuit ansatz being implemented. Thus, VQE seeks to find the optimal parameters in a fashion that is blind to control errors,  such as pulse imperfection, crosstalk and stray coupling in the device. We observe a remarkable increase in precision by using the experimentally optimal parameters rather than the theoretically optimal parameters. This finding inspires hope that VQE may provide solutions to classically intractable problems even without error correction. Additionally, these results motivate future experiments which take ``sublogical'' hardware calibration parameters, e.g. microwave pulse shapes, as variational parameters.

\begin{acknowledgments}
The authors thank Cornelius Hempel for discussions regarding VQE. J. R. M. is supported by
the Luis W. Alvarez fellowship in Computing Sciences at Lawrence Berkeley National Laboratory.
J. R. acknowledges the Air Force Office of Scientific Research for support under Award: FA9550-12-1-0046.
A. A.-G. acknowledges the Army Research Office under Award: W911NF-15-1-0256 and the Defense Security Science Engineering Fellowship managed by the Office of Naval Research under award N00014-16-1-2008. P. J. L. acknowledges the support of the National Science Foundation under grant number PHY-0955518. Devices were made at the UCSB Nanofabrication Facility, a part of the NSF-funded National Nanotechnology Infrastructure Network, and at the NanoStructures Cleanroom Facility. The authors thank H. De Raedt, M. Nocon, D. Willsch and F. Jin who brought to our attention an error in the $g_0$ values reported in an earlier version of this work.
\end{acknowledgments}

\section*{Author Contributions}
P. J. J. O. and R. Babbush contributed equally to this work. R. Babbush, H.N., A.A.-G. and J.M.M. designed the experiments. P.J.J.O. performed the experiments. J.K., R. Barends and A.M. fabricated the device. I.D.K., J.R., J.R.M., A.T., N.D., P.V.C. and P.J.L. helped R. Babbush and P.J.J.O. to compile quantum software and analyze data. R. Babbush, P.J.J.O., and J.M.M. co-wrote the manuscript. All other authors contributed to the fabrication process, experimental set-up and manuscript preparation.

\section*{Author Information}
Correspondence and requests for materials should be addressed to P.J.J.O. (pomalley@google.com), R. Babbush (babbush@google.com), or J.M.M. (jmartinis@google.com).\\

\bibliography{library}

\appendix
\section{The electronic structure problem}\label{app:ham}

The central problem of quantum chemistry is to compute the lowest energy eigenvalue of the molecular electronic structure Hamiltonian.  The eigenstates of this Hamiltonian determine almost all of the properties of interest in a molecule or material, and as the gap between the ground and first electronically excited state is often much larger than the thermal energy at room temperature, the ground state is of particular interest.  To arrive at the standard form of this Hamiltonian used in quantum computation, one begins from a collection of nuclear charges $Z_i$ and a number of electrons in the system $N$ for which the corresponding Hamiltonian is written
\begin{align}
H &= - \sum_i \frac{\nabla_{R_i}^2 }{2M_i} - \sum_i \frac{\nabla_{r_i}^2}{2} - \sum_{i,j} \frac{Z_i}{|R_i - r_j|} \notag \\
&+ \sum_{i, j > i} \frac{Z_i Z_j}{|R_i - R_j|} + \sum_{i, j>i} \frac{1}{|r_i - r_j|}
\end{align} 
 where the positions, masses, and charges of the nuclei are $R_i, M_i, Z_i$, and the positions of the electrons are $r_i$.  Here, the Hamiltonian is in atomic units of energy known as Hartree. One Hartree is $\frac{\hbar^2}{m_e e^2 a_0^2}$ (630 kcal/mol or 27.2 eV) where $m_e$, $e$ and $a_0$ denote the mass of an electron, charge of an electron and Bohr radius, respectively.

This form of the Hamiltonian and its real-space discretization are often referred to as the first quantized formulation of quantum chemistry.  Several approaches have been developed for treating this form of the problem on a quantum computer~\cite{Kassal2010}; however, the focus of this work is the second quantized formulation.  To reach the second quantized formulation, one typically first approximates the nuclei as fixed classical point charges under the Born-Oppenhemier approximation, chooses a basis $\phi_i$ in which to represent the wavefunction, and enforces anti-symmetry with the fermion creation and annihilation operators $a_i^\dagger$ and $a_j$ to give
\begin{align}
H = \sum_{pq} h_{pq} a^{\dagger}_p a_q + \frac{1}{2} \sum_{pqrs} h_{pqrs} a^{\dagger}_p a^{\dagger}_q a_r a_s
\end{align}
with 
\begin{align}
\label{eq:ints}
h_{pq} &= \int \ d\sigma \ \phi_p^*(\sigma) \left(\frac{\nabla_r^2}{2} - \sum_i \frac{Z_i}{|R_i - r|} \right)\phi_q(\sigma)  \\
h_{pqrs} &= \int \ d\sigma_1 \ d\sigma_2 \frac{\phi_p^*(\sigma_1)\phi_q^*(\sigma_2) \phi_s(\sigma_1)\phi_r(\sigma_2)}{|r_1 - r_2|}
\end{align}
where $\sigma_i$ is now a spatial and spin coordinate with $\sigma_i = (r_i, s_i)$, and the standard anti-commutation relations that determine the action of $a_i^\dagger$ and $a_j$ are $\{a_i^\dagger, a_j \} = \delta_{ij}$ and $\{a_i^\dagger, a_j^\dagger\} = \{a_i, a_j \} = 0$.  Finally, the second quantized Hamiltonian must be mapped into qubits for implementation on a quantum device.  The most common mappings used for this purpose are the Jordan-Wigner transformation \cite{Somma2002} and the Bravyi-Kitaev transformation \cite{Bravyi2002,Seeley2012,Tranter2015}.

\begin{figure}[ht]
\begin{center}
\centerline{\includegraphics[width=0.5\textwidth]{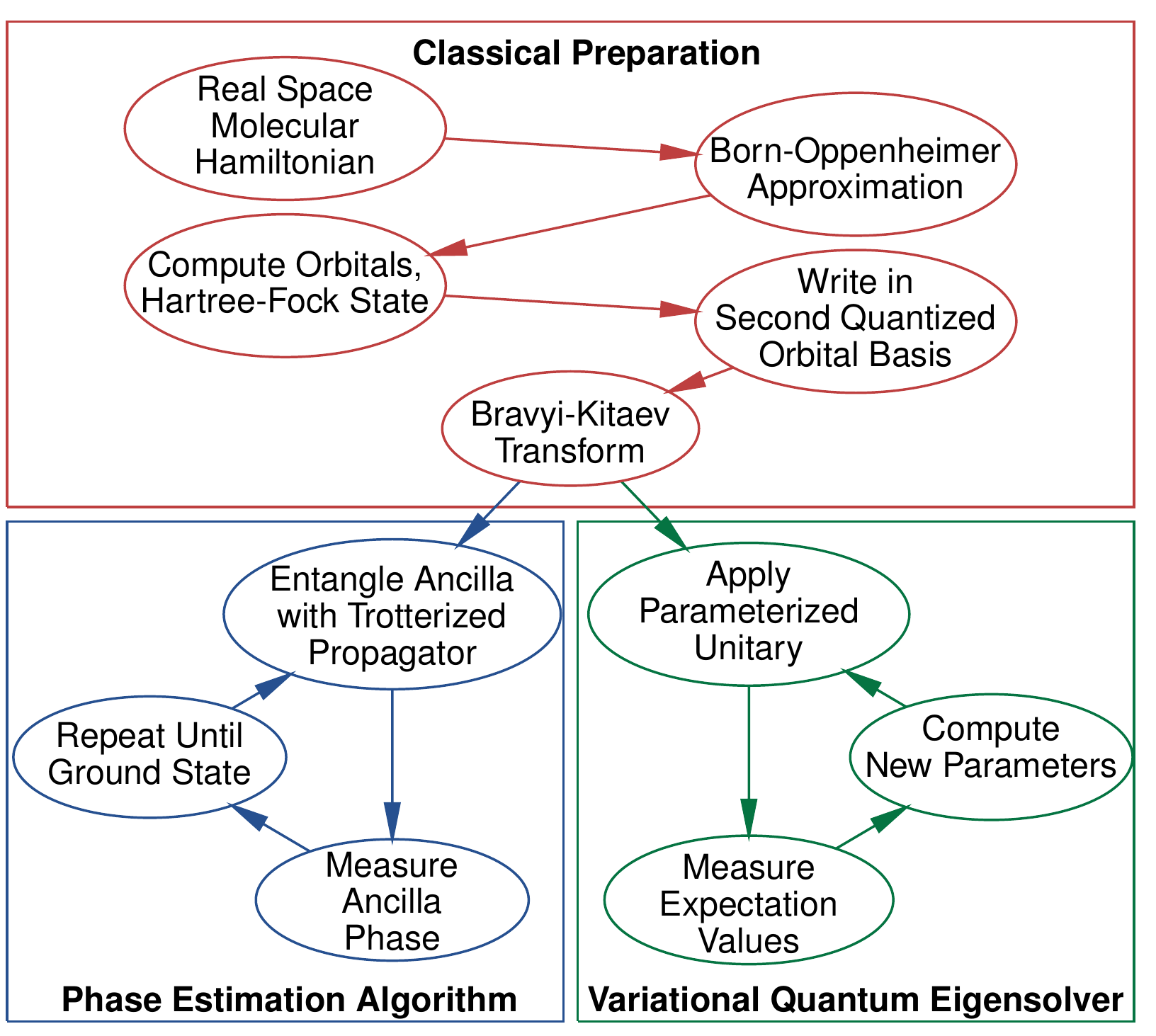}}
\caption{\label{fig:flow} A flowchart describing steps required to quantum compute molecular energies using both PEA and VQE.}
\end{center}
\end{figure}

Using the Bravyi-Kitaev transformation, the spin Hamiltonian for molecular hydrogen in the minimal (STO-6G) basis, as reported in \cite{Seeley2012}, is given by
\begin{align}
\label{eq:H2_big}
& H = f_0 \mathds{1} + f_1 Z_0 + f_2 Z_1 + f_3Z_2 +  f_1 Z_0   Z_1\\
&+ f_4 Z_0  Z_2 + f_5 Z_1   Z_3 + f_6 \textcolor{red}{X_0}   Z_1   \textcolor{red}{X
_2}  + f_6\textcolor{red}{Y_0}  Z_1  \textcolor{red}{ Y_2} \nonumber\\
& + f_7 Z_0   Z_1   Z_2 + f_4 Z_0   Z_2   Z_3 + f_3 Z_1   Z_2   Z_3\nonumber\\
&   + f_6 \textcolor{red}{X_0}   Z_1  \textcolor{red}{ X_2}   Z_3+ f_6 \textcolor{red}{Y_0}   Z_1  \textcolor{red}{ Y_2}   Z_3 + f_7 Z_0   Z_1  Z_2  Z_3\nonumber
\end{align}
where the values $\{f_i\}$ depend on the fixed bond length of the molecule. We notice that this Hamiltonian acts off-diagonally on only two qubits (the ones having tensor factors of 0 and 2), those colored in red in \eq{H2_big}. Because we start our simulations in the Hartree-Fock state, a classical basis state, we see that the Hamiltonian stabilizes qubits 1 and 3 so that they are never flipped throughout the simulation. We can use this symmetry to scalably reduce the Hamiltonian of interest to the following effective Hamiltonian which acts only on two qubits,
\begin{equation}
\widetilde{H} = g_0 \mathds{1} + g_1 Z_0 + g_2 Z_1 +  g_3 Z_0Z_1 + g_4 X_0 X_1 + g_5 Y_0 Y_1
\end{equation}
where the values $\{g_i\}$ depend on the fixed bond length of the molecule. We further note that the term $Z_0 Z_1$ commutes with all other terms in the Hamiltonian. Since the ground state of the total Hamiltonian certainly has support on the Hartree-Fock state, we know the contribution to the total energy of $Z_0 Z_1$ (it is given by the expectation of those terms with the Hartree-Fock state). Steps to prepare this Hamiltonian are summarized in the upper-half of \fig{flow}.

\section{Experimental methods for VQE}\label{app:vqe}

For the VQE experiment, the qubits $q_0$ and $q_1$ are used, at 4.49 and 5.53 GHz, respectively, while all the other qubits are detuned to 3 GHz and below. $X_\pi$, $Y_\pi$, $\pm X_{\pi/2}$, and $\pm Y_{\pi/2}$ gates are 25 ns long, and pulse amplitudes and detunings from $f_{10}$ are optimized with ORBIT; for these parameters, additional pulse shaping (e.g. DRAG) proved unnecessary (see \cite{Barends2016} for details of pulse detuning and shaping). The amplitude, trajectory, and compensating single-qubit phases of the CZ$_\pi$ gate are optimized with ORBIT as well. The duration of the CZ$_\pi$ is 55 ns, during which the frequency of $q_0$ is fixed and $q_1$ is moved. The rotation $Z_\theta$ (the adjustable parameter in \eq{xy}) is implemented as a phase shift on all subsequent gates. As operated here, $q_0$ and $q_1$ have energy relaxation times $T_1 = 62.8$ and $21.4\,\mu s$, and Ramsey decay times $T_2^* = 1.1$ and $1.9\,\mu s$, respectively.

The expectation values used to calculate the energy of the prepared state are measured with partial tomography; for example, $X_1X_0$ is measured by applying $Y_{\pi/2}$ gates to each qubit prior to measurement. We emphasize that for chemistry problems, the number of measurements scales polynomially \cite{McClean2015}. Readout duration is set to 1000 ns for higher fidelity (compared to \cite{Kelly2015}, where the ``measure''/odd-numbered qubits utilized much shorter readout). In addition to discriminating between $\ket{0}$ and $\ket{1}$, higher level qubit states were also measured (called $\ket{2}$ for simplicity). Readout fidelities are typically $>$99\% for $\ket{0}$, and $\sim$95\% for $\ket{1}$ and $\ket{2}$, and measurement probabilities are corrected for readout error. After readout correction, experiments where one of the qubits is measured in $\ket{2}$ are dropped; any probability to be in $\ket{2}$ is set to zero and remaining probabilities are renormalized.

\begin{figure}[ht]
\begin{center}
\centerline{\includegraphics[width=0.5\textwidth]{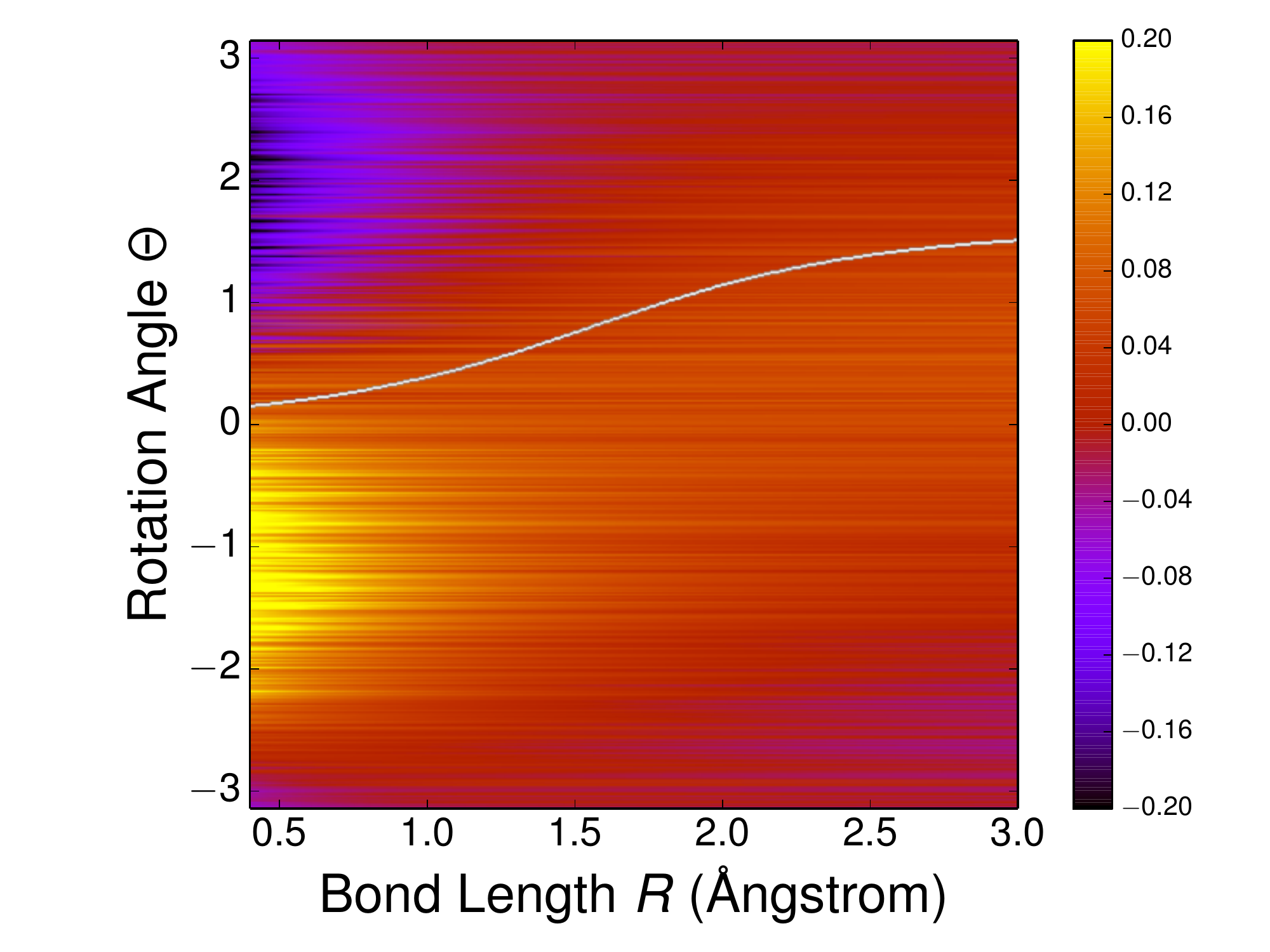}}
\caption{\label{fig:error_surface} Errors in the VQE energy surface (in Hartree) as a function of bond length and rotation angle. This plot looks somewhat like the derivative of \figa{data}{b} with respect to $R$ and $\theta$ because errors are greatest where the energy is most sensitive to changes in system parameters. As in \figa{data}{b}, the white curve traces the theoretical minimum energy which is seen to be in good agreement with the data. Note that while errors in the energy surface are sometimes negative, all energies are bounded from below by the variational minimum.}
\end{center}
\end{figure}

The circuit pulse sequence used to implement the UCC sequence in \eq{xy} is shown in \fig{vqe}. The experiment is performed in different gauges of the Bravyi-Kitaev transform; these correspond to the $\ket{0}$ ($\ket{1}$) state of $q_0$ representing the first orbital being unoccupied (occupied) or occupied (unoccupied), and similarly for $q_1$ representing the parity of the first two orbitals being even (odd) or odd (even). In practice, a gauge change means a flip of the value of one or both qubits in the Hartree-Fock (HF) input state, and a sign change on the relevant terms of the Hamiltonian. In the standard gauge, the HF state is $\ket{01}$ and is prepared with an $X_\pi$ gate on $q_0$. Statistics from the experiment in these gauges are then averaged together. We also drop the first $-Y_{\pi/2}$ on $q_0$; for an input state of either $\ket{0}$ or $\ket{1}$, it has no effect given that $X_{\pi/2}$ is the only gate preceding it.

The energy for a given nuclear separation $R$ is computed by calculating the value of the Hamiltonian with the expectation values measured for each $\theta$ and choosing the smallest energy. This is done for all values of $R$ to construct the energy surface. \figa{data}{a} shows the raw expectation values (after readout correction); \figa{data}{b} shows the measured energy versus $\theta$ for each value of $R$ and \fig{error_surface} shows the errors in that surface. Error bars were computed from a Gaussian process regression \cite{Bishop2006} applied to the potential energy curve obtained from \figa{data}{b} using error estimates propagated from the shot-noise limited measurements shown in \figa{data}{a}.

\section{Experimental methods for PEA}\label{app:pea}

The PEA experiment uses three qubits: $q_0$ for the ancilla, and $q_1$ and $q_2$ for the register. Operating frequencies are 4.56, 5.65, and 4.80 GHz for $q_0$, $q_1$, and $q_2$, respectively. Pulse tune-up is the same as for the VQE experiment. For the entangling gates (CZ$_\phi$ between $q_0$ and $q_1$, and CZ$_\pi$ between $q_1$ and $q_2$), however, the adjacent non-interacting qubit must be decoupled from the interaction. For the CZ$_\pi$, $q_0$ is decoupled with paired $X_\pi$ and $-X_\pi$ pulses; this has the effect of ``echoing out'' any acquired state-dependent phase on $q_0$ from $q_1$ and vice versa, while minimizing stray single-qubit phases on $q_0$ by keeping its frequency stationary. For the CZ$_\phi$, however, $q_2$ is detuned to frequencies significantly below the $q_0$-$q_1$ interaction; while this makes single-qubit phases on $q_2$ harder to compensate, it is more effective at minimizing the impact of $q_2$ on the CZ$_\phi$ gate. This combination of decoupling methods was found to be optimal to minimize error on the phase of $q_0$, which is the critical parameter in the PEA experiment.

As the CZ$_\phi$ gate varies the amplitude of $q_1$'s frequency trajectory over a wide range (approximately 200 MHz to 950 MHz) particular values of $\phi$ can be more sensitive lossy parts of the $q_1$'s frequency spectrum that are rapidly swept past and easily compensated for in the standard case of only tuning up $\phi = \pi$. Therefore, for some values of $\phi$ it is necessary to individually tune in compensating phases on $q_0$. This is implemented by executing the individual term of the Hamiltonian, varying the compensating phase on $q_0$, and fitting for the value that minimizes the error of that term. After performing this careful compensation when necessary, the experiment produces the bit values (0 or 1) for each different Hamiltonian (i.e. each separation $R$) at each evolution time $t$ that match those predicted by numerical simulation.

\begin{figure}[t]
\begin{center}
\centerline{\includegraphics[width=0.5\textwidth]{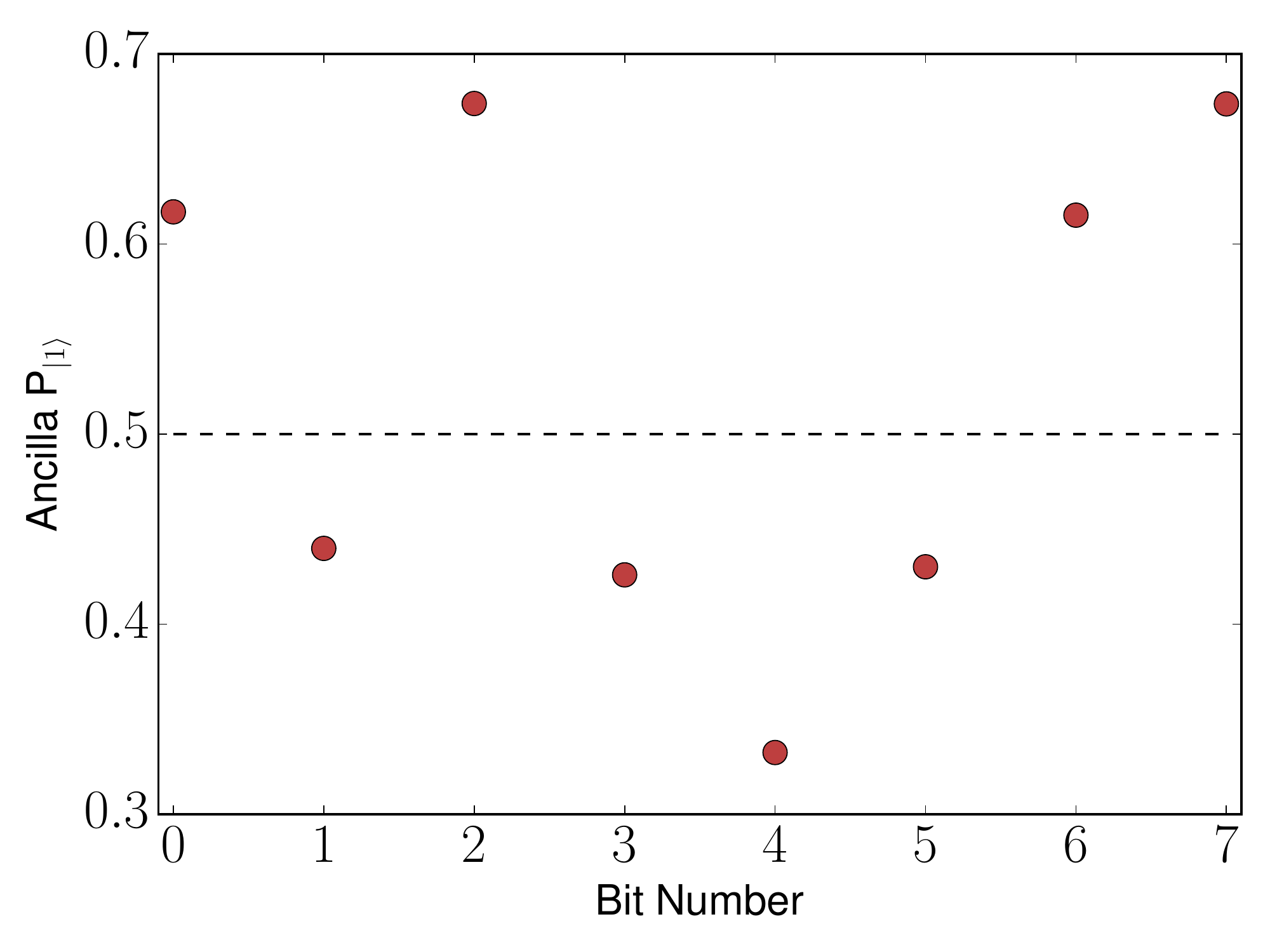}}
\caption{\label{fig:sample_pea} Example data for a single PEA experiment, run at $R=1.55$ \AA.
The results are shown without phase kickback from the measurements of the previous bit.
The line at $P_{\ket{1}} = 0.5$ discriminates a measurement of 1 from 0.}
\end{center}
\end{figure}

As operated in this experiment, $q_0$, $q_1$, and $q_2$ have $T_1$ values of 48.1, 23.7, and 43.0 $\mu$s, and $T_2^*$ times of 1.3, 1.6, and 0.8 $\mu$s, respectively. \fig{pea} shows the pulses for one iteration of the PEA experiment; \fig{sample_pea} shows an example of the measurement results for one value of $R$. The parameters at each $R$ are given in \tab{pea_table}. For reference, included in this section are the implementations of all the terms of the Bravyi-Kitaev Hamiltonian for molecular hydrogen. In the following diagrams, $\alpha$ is the ancilla qubit ($q_0$ in the experiment), and 0 and 1 are the register qubits ($q_1$ and $q_2$ in the experiment). We must always be aware that representing our terms in terms of these gates, and then in terms of the actual basis, is not necessarily the most efficient approach.

\subsection*{CNOT}

CNOT is implemented as a CZ$_\pi$ and two rotations.

\centerline{
\Qcircuit @C=1em @R=.7em {
& \ctrl{1}  & \qw & & & \qw & \ctrl{1} & \qw & \qw \\
& \targ & \qw & \push{\rule{.3em}{0em}=\rule{.3em}{0em}}& & \gate{-Y_{\pi/2}} & \ctrl{-1} & \gate{Y_{\pi/2}} & \qw \\
}
}

\subsection*{SWAP}

\noindent SWAP is implemented as three consecutive CZ$_\pi$ gates with intermediate rotations. \\

\centerline{
\Qcircuit @C=.7em @R=.7em {
& \qswap  & \qw & & & \qw & \ctrl{1} & \gate{-Y_{\pi/2}} & \ctrl{1} & \gate{Y_{\pi/2}} & \ctrl{1} &\qw & \qw \\
& \qswap\qwx & \qw & \push{\rule{.3em}{0em}=\rule{.3em}{0em}}& & \gate{-Y_{\pi/2}} & \ctrl{-1} & \gate{Y_{\pi/2}} & \ctrl{-1} & \gate{-Y_{\pi/2}} & \ctrl{-1} & \gate{Y_{\pi/2}} & \qw \\
}
}

\subsection*{Controlled evolution under $Z_0$}
\noindent $Z_0$ is implemented as CZ$_{\phi}$ and a $z$ rotation on the control qubit. \\

\centerline{
\Qcircuit @C=1em @R=1.5em {
q_\alpha &&\gate{-Z_{\theta/2}} & \ctrl{1} & \qw \\
q_0 &&\qw& \gate{Z_{\theta}} & \qw \\
}
}

\subsection*{Controlled evolution under $Z_1$}
\noindent $Z_1$ is the same as $Z_0$, but surrounded by SWAP gates so that the ancilla interacts with the other qubit.\\

\centerline{
\Qcircuit @C=1em @R=1.5em {
q_\alpha & & \gate{-Z_{\theta/2}} & \ctrl{1} & \qw & \qw \\
q_0 && \qswap & \gate{Z_{\theta}} & \qswap & \qw \\
q_1 && \qswap\qwx & \qw & \qswap\qwx & \qw \\\\
}
}

\subsection*{Controlled evolution under $X_0 X_1$}
\noindent For $X_0 X_1$, we first change bases with $Y_{\pi/2}$ gates, then compute the parity of the register qubits with a CNOT, then apply the controlled phase, and finally undo the parity computation and basis change. Note that the $Y_{\pi/2}$ gates will cancel on the middle qubit with our CNOT implementation. \\

\centerline{
\Qcircuit @C=1em @R=1.5em {
q_\alpha && \gate{-Z_{\theta/2}} & \qw & \ctrl{1} & \qw & \qw & \qw \\
q_0 && \gate{Y_{\pi/2}} & \targ & \gate{Z_\theta} & \targ & \gate{-Y_{\pi/2}} & \qw \\
q_1 && \gate{Y_{\pi/2}} & \ctrl{-1} & \qw & \ctrl{-1} & \gate{-Y_{\pi/2}} & \qw \\\\
}
}
\subsection*{Controlled evolution under $Y_0 Y_1$}
\noindent $Y_0 Y_1$ is the same as $X_0 X_1$ with a different basis change. \\

\centerline{
\Qcircuit @C=1em @R=1.5em {
q_\alpha && \gate{-Z_{\theta/2}} & \qw & \ctrl{1} & \qw & \qw & \qw \\
q_0 && \gate{-X_{\pi/2}} & \targ & \gate{Z_\theta} & \targ & \gate{X_{\pi/2}} & \qw \\
q_1 && \gate{-X_{\pi/2}} & \ctrl{-1} & \qw & \ctrl{-1} & \gate{X_{\pi/2}} & \qw \\\\
}
}

\section{Unitary coupled cluster}\label{app:ucc}

The application of VQE requires the choice of an ansatz, and in this work we have focused on the unitary coupled cluster (UCC) ansatz.  This ansatz is a unitary variant of the method sometimes referred to as the ``gold standard of quantum chemistry'', namely coupled cluster with single and double excitations with perturbative triples excitations \cite{Helgaker2002}. The unitary variant has the advantage of satisfying a variational principle with respect to all possible parameterizations \cite{Hoffmann1988,Bartlett1989,Taube2006}.  Furthermore, UCC can be easily applied to a multireference initial state whereas one of the major shortcomings of traditional coupled cluster is that it can only be applied to a single Slater determinant \cite{Hoffmann1988,Bartlett1989,Taube2006}. While the unitary variant has no efficient preparation scheme on a classical computer, scalable methods of preparation for a fixed set of parameters on a quantum device have now been documented several times \cite{Peruzzo2013,Yung2013,Wecker2015,McClean2015}.

The UCC ansatz $\ket{\varphi (\vec \theta)}$ is defined with respect to a reference state, which in this work we take to be the Hartree-Fock state $\ket \phi$,
\begin{align}
 \ket{\varphi(\vec \theta)} = U(\vec \theta) \ket{\varphi} = e^{T(\vec \theta) - T(\vec \theta)^\dagger} \ket{\phi}
\end{align}
where $T(\vec \theta)$ is the anti-Hermitian cluster operator:
\begin{align}
T & = \sum_k {}^{(k)}T(\vec \theta)\\
{}^{(1)} T(\vec \theta) & = \sum_{\substack{i_1 \in \text{occ} \\ a_1 \in \text{virt}}} \theta^{a_1}_{i_1} a_{a_1}^\dagger a_{i_1} \\
{}^{(2)} T(\vec \theta) & = \frac{1}{4} \sum_{\substack{i_1,i_2 \in \text{occ}\\ a_1,a_2 \in \text{virt}}} \theta^{a_1,a_2}_{i_1,i_2} a_{a_2}^\dagger a_{i_2} a_{a_1}^\dagger a_{i_1}
\end{align}
where the occ and virt spaces are defined as the occupied and unoccupied sites in the Hartree-Fock state and the definition of higher-order cluster operators ${}^{(k)}T$ follows naturally.  When only including up to the first two terms in the cluster expansion, we term the ansatz unitary coupled cluster with single and doubles excitations (UCCSD) \cite{Helgaker2002}.

The task within VQE is to determine the optimal values of the one- and two-body cluster amplitudes $\theta^{a_1}_{i_1}$ and $\theta^{a_1, a_2}_{i_1, i_2}$, which are determined by the variational minimum of a nonlinear function. As with all nonlinear minimizations, the choice of starting parameters is key to algorithmic performance.  As in classical coupled cluster, we can determine the starting amplitudes perturbatively through M{\"o}ller-Plesset perturbation theory (MP2) \cite{Helgaker2002}. For molecular hydrogen in the minimal basis, there is exactly one term in the UCCSD ansatz.

The MP2 guess amplitudes are given by the equations
\begin{align}
 \theta^{a}_{i} = 0, \quad \quad \quad  \theta^{ab}_{ij} = \frac{h_{ijba} - h_{ijab}}{\epsilon_i + \epsilon_j - \epsilon_a - \epsilon_b}
\end{align}
where $\epsilon_a$ refer to the 1-electron occupied and virtual orbital energies from the Hartree-Fock calculation and the $h_{ijab}$ are computed as in \eq{ints}. In the MP2 guess, the vanishing of the singles amplitudes is a result of the fact that single excitations away from the Hartree-Fock reference do not couple through the Hamiltonian as a consequence of Brillouin's theorem \cite{Helgaker2002}.  As the solution of the classical coupled cluster equations is also efficient, it is possible to use amplitudes from a method like CCSD as starting values as well.  We note in both cases however, that the single-reference, perturbative nature of these constructions may lead to poor initial guesses for systems with strong multireference character or entanglement.  In these cases the amplitudes may represent poor guesses, requiring more iterations for convergence.  As such, a better initial guess in such problems may be a related optimization, such as a different molecular geometry of the same system.  In cases where the perturbative estimates are accurate, one can discard operations related to very small amplitudes in the state preparation circuit, leading to computational savings.

\begin{table*}[p]
\centering
\vspace{-.75cm}
\caption{\label{tab:pea_table} The Hamiltonian coefficients for \eq{H2} and parameters for the PEA experiment for each value of $R$.}
\begin{tabular}{c|c|c|c|c|c|c|c|c|c}
$R$ & \openone & $Z_0$ & $Z_1$ & $Z_0Z_1$ & $X_0 X_1$ & $Y_0 Y_1$ & $t_0$ & Ordering & Trotter Error \\ \hline
0.20 & 2.8489 & 0.5678 & -1.4508 & 0.6799 & 0.0791 & 0.0791 & 1.500 & $Z_0 \cdot X_0 X_1 \cdot Z_1 \cdot Y_0 Y_1$ & 0.0124 \\
0.25 & 2.1868 & 0.5449 & -1.2870 & 0.6719 & 0.0798 & 0.0798 & 1.590 & $Z_0 \cdot Y_0 Y_1 \cdot Z_1 \cdot X_0 X_1$ & 0.0521 \\
0.30 & 1.7252 & 0.5215 & -1.1458 & 0.6631 & 0.0806 & 0.0806 & 1.770 & $X_0 X_1 \cdot Z_0 \cdot Y_0 Y_1 \cdot Z_1$ & 0.0111 \\
0.35 & 1.3827 & 0.4982 & -1.0226 & 0.6537 & 0.0815 & 0.0815 & 2.080 & $Z_0 \cdot X_0 X_1 \cdot Z_1 \cdot Y_0 Y_1$ & 0.0368 \\
0.40 & 1.1182 & 0.4754 & -0.9145 & 0.6438 & 0.0825 & 0.0825 & 2.100 & $Z_0 \cdot X_0 X_1 \cdot Z_1 \cdot Y_0 Y_1$ & 0.0088 \\
0.45 & 0.9083 & 0.4534 & -0.8194 & 0.6336 & 0.0835 & 0.0835 & 2.310 & $X_0 X_1 \cdot Z_0 \cdot Y_0 Y_1 \cdot Z_1$ & 0.0141 \\
0.50 & 0.7381 & 0.4325 & -0.7355 & 0.6233 & 0.0846 & 0.0846 & 2.580 & $Z_0 \cdot X_0 X_1 \cdot Z_1 \cdot Y_0 Y_1$ & 0.0672 \\
0.55 & 0.5979 & 0.4125 & -0.6612 & 0.6129 & 0.0858 & 0.0858 & 2.700 & $Z_0 \cdot X_0 X_1 \cdot Z_1 \cdot Y_0 Y_1$ & 0.0147 \\
0.60 & 0.4808 & 0.3937 & -0.5950 & 0.6025 & 0.0870 & 0.0870 & 2.250 & $Z_0 \cdot X_0 X_1 \cdot Z_1 \cdot Y_0 Y_1$ & 0.0167 \\
0.65 & 0.3819 & 0.3760 & -0.5358 & 0.5921 & 0.0883 & 0.0883 & 3.340 & $Z_1 \cdot X_0 X_1 \cdot Z_0 \cdot Y_0 Y_1$ & 0.0175 \\
0.70 & 0.2976 & 0.3593 & -0.4826 & 0.5818 & 0.0896 & 0.0896 & 0.640 & $Z_0 \cdot Y_0 Y_1 \cdot Z_1 \cdot X_0 X_1$ & 0.0171 \\
0.75 & 0.2252 & 0.3435 & -0.4347 & 0.5716 & 0.0910 & 0.0910 & 0.740 & $Z_0 \cdot Y_0 Y_1 \cdot Z_1 \cdot X_0 X_1$ & 0.0199 \\
0.80 & 0.1626 & 0.3288 & -0.3915 & 0.5616 & 0.0925 & 0.0925 & 0.790 & $Z_0 \cdot Y_0 Y_1 \cdot Z_1 \cdot X_0 X_1$ & 0.0291 \\
0.85 & 0.1083 & 0.3149 & -0.3523 & 0.5518 & 0.0939 & 0.0939 & 3.510 & $Z_0 \cdot X_0 X_1 \cdot Z_1 \cdot Y_0 Y_1$ & 0.0254 \\
0.90 & 0.0609 & 0.3018 & -0.3168 & 0.5421 & 0.0954 & 0.0954 & 3.330 & $Z_0 \cdot X_0 X_1 \cdot Z_1 \cdot Y_0 Y_1$ & 0.0283 \\
0.95 & 0.0193 & 0.2895 & -0.2845 & 0.5327 & 0.0970 & 0.0970 & 4.090 & $X_0 X_1 \cdot Z_0 \cdot Y_0 Y_1 \cdot Z_1$ & 0.0328 \\
1.00 & -0.0172 & 0.2779 & -0.2550 & 0.5235 & 0.0986 & 0.0986 & 4.360 & $Z_0 \cdot X_0 X_1 \cdot Z_1 \cdot Y_0 Y_1$ & 0.0362 \\
1.05 & -0.0493 & 0.2669 & -0.2282 & 0.5146 & 0.1002 & 0.1002 & 4.650 & $Z_1 \cdot X_0 X_1 \cdot Z_0 \cdot Y_0 Y_1$ & 0.0405 \\
1.10 & -0.0778 & 0.2565 & -0.2036 & 0.5059 & 0.1018 & 0.1018 & 4.280 & $Z_1 \cdot X_0 X_1 \cdot Z_0 \cdot Y_0 Y_1$ & 0.0243 \\
1.15 & -0.1029 & 0.2467 & -0.1810 & 0.4974 & 0.1034 & 0.1034 & 5.510 & $Z_0 \cdot X_0 X_1 \cdot Z_1 \cdot Y_0 Y_1$ & 0.0497 \\
1.20 & -0.1253 & 0.2374 & -0.1603 & 0.4892 & 0.1050 & 0.1050 & 5.950 & $Z_0 \cdot Y_0 Y_1 \cdot Z_1 \cdot X_0 X_1$ & 0.0559 \\
1.25 & -0.1452 & 0.2286 & -0.1413 & 0.4812 & 0.1067 & 0.1067 & 6.360 & $X_0 X_1 \cdot Z_1 \cdot Y_0 Y_1 \cdot Z_0$ & 0.0585 \\
1.30 & -0.1629 & 0.2203 & -0.1238 & 0.4735 & 0.1083 & 0.1083 & 0.660 & $Z_1 \cdot X_0 X_1 \cdot Z_0 \cdot Y_0 Y_1$ & 0.0905 \\
1.35 & -0.1786 & 0.2123 & -0.1077 & 0.4660 & 0.1100 & 0.1100 & 9.810 & $Z_0 \cdot X_0 X_1 \cdot Z_1 \cdot Y_0 Y_1$ & 0.0694 \\
1.40 & -0.1927 & 0.2048 & -0.0929 & 0.4588 & 0.1116 & 0.1116 & 9.930 & $Z_0 \cdot X_0 X_1 \cdot Z_1 \cdot Y_0 Y_1$ & 0.0755 \\
1.45 & -0.2053 & 0.1976 & -0.0792 & 0.4518 & 0.1133 & 0.1133 & 5.680 & $Y_0 Y_1 \cdot Z_0 \cdot X_0 X_1 \cdot Z_1$ & 0.0142 \\
1.50 & -0.2165 & 0.1908 & -0.0666 & 0.4451 & 0.1149 & 0.1149 & 10.200 & $Z_1 \cdot X_0 X_1 \cdot Z_0 \cdot Y_0 Y_1$ & 0.0885 \\
1.55 & -0.2265 & 0.1843 & -0.0549 & 0.4386 & 0.1165 & 0.1165 & 9.830 & $Z_0 \cdot X_0 X_1 \cdot Z_1 \cdot Y_0 Y_1$ & 0.0917 \\
1.60 & -0.2355 & 0.1782 & -0.0442 & 0.4323 & 0.1181 & 0.1181 & 8.150 & $Z_0 \cdot Y_0 Y_1 \cdot Z_1 \cdot X_0 X_1$ & 0.0416 \\
1.65 & -0.2436 & 0.1723 & -0.0342 & 0.4262 & 0.1196 & 0.1196 & 8.240 & $X_0 X_1 \cdot Z_0 \cdot Y_0 Y_1 \cdot Z_1$ & 0.0488 \\
1.70 & -0.2508 & 0.1667 & -0.0251 & 0.4204 & 0.1211 & 0.1211 & 0.520 & $Z_1 \cdot X_0 X_1 \cdot Z_0 \cdot Y_0 Y_1$ & 0.0450 \\
1.75 & -0.2573 & 0.1615 & -0.0166 & 0.4148 & 0.1226 & 0.1226 & 0.520 & $Z_0 \cdot Y_0 Y_1 \cdot Z_1 \cdot X_0 X_1$ & 0.0509 \\
1.80 & -0.2632 & 0.1565 & -0.0088 & 0.4094 & 0.1241 & 0.1241 & 1.010 & $Z_0 \cdot X_0 X_1 \cdot Z_1 \cdot Y_0 Y_1$ & 0.0663 \\
1.85 & -0.2684 & 0.1517 & -0.0015 & 0.4042 & 0.1256 & 0.1256 & 0.530 & $Z_1 \cdot X_0 X_1 \cdot Z_0 \cdot Y_0 Y_1$ & 0.0163 \\
1.90 & -0.2731 & 0.1472 & 0.0052 & 0.3992 & 0.1270 & 0.1270 & 1.090 & $X_0 X_1 \cdot Z_0 \cdot Z_1 \cdot Y_0 Y_1$ & 0.0017 \\
1.95 & -0.2774 & 0.1430 & 0.0114 & 0.3944 & 0.1284 & 0.1284 & 0.610 & $X_0 X_1 \cdot Z_1 \cdot Z_0 \cdot Y_0 Y_1$ & 0.0873 \\
2.00 & -0.2812 & 0.1390 & 0.0171 & 0.3898 & 0.1297 & 0.1297 & 1.950 & $Z_1 \cdot Z_0 \cdot X_0 X_1 \cdot Y_0 Y_1$ & 0.0784 \\
2.05 & -0.2847 & 0.1352 & 0.0223 & 0.3853 & 0.1310 & 0.1310 & 4.830 & $X_0 X_1 \cdot Y_0 Y_1 \cdot Z_0 \cdot Z_1$ & 0.0947 \\
2.10 & -0.2879 & 0.1316 & 0.0272 & 0.3811 & 0.1323 & 0.1323 & 1.690 & $Y_0 Y_1 \cdot X_0 X_1 \cdot Z_0 \cdot Z_1$ & 0.0206 \\
2.15 & -0.2908 & 0.1282 & 0.0317 & 0.3769 & 0.1335 & 0.1335 & 0.430 & $X_0 X_1 \cdot Y_0 Y_1 \cdot Z_0 \cdot Z_1$ & 0.0014 \\
2.20 & -0.2934 & 0.1251 & 0.0359 & 0.3730 & 0.1347 & 0.1347 & 1.750 & $Z_0 \cdot Z_1 \cdot X_0 X_1 \cdot Y_0 Y_1$ & 0.0107 \\
2.25 & -0.2958 & 0.1221 & 0.0397 & 0.3692 & 0.1359 & 0.1359 & 11.500 & $X_0 X_1 \cdot Z_1 \cdot Z_0 \cdot Y_0 Y_1$ & 0.0946 \\
2.30 & -0.2980 & 0.1193 & 0.0432 & 0.3655 & 0.1370 & 0.1370 & 0.420 & $Z_0 \cdot Z_1 \cdot X_0 X_1 \cdot Y_0 Y_1$ & 0.0370 \\
2.35 & -0.3000 & 0.1167 & 0.0465 & 0.3620 & 0.1381 & 0.1381 & 0.470 & $Z_1 \cdot Z_0 \cdot Y_0 Y_1 \cdot X_0 X_1$ & 0.0762 \\
2.40 & -0.3018 & 0.1142 & 0.0495 & 0.3586 & 0.1392 & 0.1392 & 10.100 & $X_0 X_1 \cdot Z_1 \cdot Z_0 \cdot Y_0 Y_1$ & 0.0334 \\
2.45 & -0.3035 & 0.1119 & 0.0523 & 0.3553 & 0.1402 & 0.1402 & 11.200 & $Z_0 \cdot Z_1 \cdot X_0 X_1 \cdot Y_0 Y_1$ & 0.0663 \\
2.50 & -0.3051 & 0.1098 & 0.0549 & 0.3521 & 0.1412 & 0.1412 & 0.580 & $Z_0 \cdot Y_0 Y_1 \cdot X_0 X_1 \cdot Z_1$ & 0.0296 \\
2.55 & -0.3066 & 0.1078 & 0.0572 & 0.3491 & 0.1422 & 0.1422 & 11.000 & $Z_0 \cdot Z_1 \cdot X_0 X_1 \cdot Y_0 Y_1$ & 0.0550 \\
2.60 & -0.3079 & 0.1059 & 0.0594 & 0.3461 & 0.1432 & 0.1432 & 11.000 & $Z_0 \cdot X_0 X_1 \cdot Y_0 Y_1 \cdot Z_1$ & 0.0507 \\
2.65 & -0.3092 & 0.1042 & 0.0614 & 0.3433 & 0.1441 & 0.1441 & 11.040 & $Z_1 \cdot X_0 X_1 \cdot Y_0 Y_1 \cdot Z_0$ & 0.0490 \\
2.70 & -0.3104 & 0.1026 & 0.0632 & 0.3406 & 0.1450 & 0.1450 & 0.400 & $Z_0 \cdot Z_1 \cdot Y_0 Y_1 \cdot X_0 X_1$ & 0.0471 \\
2.75 & -0.3115 & 0.1011 & 0.0649 & 0.3379 & 0.1458 & 0.1458 & 0.450 & $Y_0 Y_1 \cdot Z_0 \cdot Z_1 \cdot X_0 X_1$ & 0.0061 \\
2.80 & -0.3125 & 0.0997 & 0.0665 & 0.3354 & 0.1467 & 0.1467 & 0.950 & $Z_0 \cdot Y_0 Y_1 \cdot X_0 X_1 \cdot Z_1$ & 0.0368 \\
2.85 & -0.3135 & 0.0984 & 0.0679 & 0.3329 & 0.1475 & 0.1475 & 10.600 & $Z_0 \cdot X_0 X_1 \cdot Y_0 Y_1 \cdot Z_1$ & 0.0324 \\
\end{tabular}

\end{table*}

\end{document}